\documentclass[12pt]{emulateapj}

\usepackage{subfigure}
\usepackage{graphics}
\usepackage{natbib}
\usepackage{hyperref}
\usepackage{longtable}

\def\kms{~km~s$^{-1}$\ }
\def\kmsc{~km~s$^{-1}$}

\def\arcm{\char'023\ }
\def\arcmn{\char'023 }

\def\hub{\ifmmode H_\circ\else H$_\circ$\fi}
\def\lsim{\mathop{\hbox{${\lower3.8pt\hbox{$<$}}\atop{\raise0.2pt\hbox{$\sim$}}$}}}
\def\gsim{\mathop{\hbox{${\lower3.8pt\hbox{$>$}}\atop{\raise0.2pt\hbox{$\sim$}}$}}}

\shorttitle{M31 Globular Clusters}
\shortauthors{Caldwell \& Romanowsky}

\begin{document}

\title{Star Clusters in M31: VII. Global Kinematics and Metallicity Subpopulations of the Globular Clusters}

\author{Nelson Caldwell} 
\affil{Harvard-Smithsonian Center for Astrophysics, 60 Garden Street, Cambridge, MA 02138, USA
\\ electronic mail: caldwell@cfa.harvard.edu}

\author{Aaron J. Romanowsky} 
\affil{Department of Physics \& Astronomy, San Jos\'e State University,
One Washington Square, San Jose, CA 95192, USA \\ 
University of California Observatories, 1156 High Street, Santa Cruz, CA 95064, USA\\
electronic mail: aaron.romanowsky@sjsu.edu 
}

\shorttitle{M31 Globulars}

\begin{abstract}

We carry out a joint spatial--kinematical--metallicity analysis of globular clusters (GCs) around the Andromeda Galaxy (M31),
using a homogeneous, high-quality spectroscopic dataset.
In particular, we remove the contaminating  young clusters that have plagued many previous analyses. We find that
the clusters can be divided into three major metallicity groups based on their radial distributions: (1) an inner metal-rich group 
([Fe/H] $ > -0.4$), (2) a group with intermediate metallicity (with median [Fe/H]=$-1$), (3) and a 
metal-poor group, with [Fe/H] $< -1.5$. The metal-rich group has kinematics and spatial properties like the disk of M31,
while the two more metal-poor groups show mild prograde rotation overall, with larger dispersions --
in contrast to previous claims of stronger rotation. 
The metal-poor GCs are the least concentrated group; 
such clusters occur five times less frequently in the central bulge than do clusters of higher metallicity.
Despite some well-known differences between the M31 and Milky Way GC systems,
our revised analysis points to remarkable similarities in their chemodynamical properties, 
which could help elucidate the different formation stages of galaxies and their GCs.
In particular, the M31 results motivate further exploration of a metal-rich GC formation mode in situ,
within high-redshift, clumpy galactic disks.

\end{abstract}


\section{Introduction}\label{intro}

Globular star clusters (GCs) have a venerable history as unique tracers of the global properties and formation histories of galaxies, 
where they can be used as proxies for stellar densities, kinematics, and metallicities over vast scales.
The Milky Way (MW) is an archetypal case, where the classic work of \citet{SZ1978} inferred from GCs that the outer halo assembled
through the protracted infall and merging of smaller galaxies.
The MW GC system is currently thought to consist of two or three basic components, with each of them representing a distinct, major mode
of ancient star formation \citep{zinn85,harris01}. 
These include an inner component that is flattened, strongly rotating, metal-rich, and
identified with the stellar bulge and/or thick disk. An outer component is quasi-spherical, weakly rotating, metal-poor, and
corresponds to the stellar halo.  These halo GCs may be further divided into inner and outer components, sometimes called the ``old'' and
``young'' halo, with distinct kinematic and chemical abundance patterns \citep{mackey04,forbes10,keller12}.
 Similar divisions have been made in the stellar halo \citep{carollo07, morrison09}, with an outer halo interpreted as the debris of satellite galaxies,
  and an inner halo 
  whose origin is less certain but may have formed ``in situ.''

It is important to ascertain how general these patterns are for other galaxies, and the natural place to start is our nearest
large neighbor, the Andromeda Galaxy (M31). Here there are two broad, well-established differences: 
(1) the M31 GC system is more populous than the MW system, by a factor of $\sim$~2--3, and
(2) it does not exhibit the same obvious bimodality in metallicity  \citep{barmby00,galleti09,C11,cz13}.
Both of these aspects may be reflections of dramatic differences discovered in these galaxies' stellar halos, where the M31 halo appears
much more metal-enriched, with massive substructures suggesting a more active satellite accretion history (e.g.,  \citealt{mcconnachie09}).

A third difference has also been noted, involving the rotation.
Although early studies of the M31 GC system found relatively weak rotation, particularly among the metal-poor GCs
\citep{huchra82,freeman83,elson88,huchra91,barmby00},
later studies found stronger rotation \citep{perrett02,lee08,deason11}.
This is important because the low rotation in the MW is part of the classical line of evidence for
accretion in galaxy halos, where myriad minor mergers from quasi-random directions lead to relatively low net angular momentum
(e.g., \citealt{mccarthy}).
Hence, one natural interpretation of the more recent M31 results is that the rotation traces a past major merger that spun up the entire galaxy \citep{bekki10}.
However, observations of M31 GCs have been historically fraught with difficulty, owing to interference from its
massive, extended disk -- leading to false alarms about strong rotation patterns that were actually caused by
young disk cluster contamination in the GC sample (as seen in the \citealt{perrett02}, \citealt{morrison04}, and \citealt{beasley04} papers).

Recent observational efforts have turned to the kinematics of GCs in the outer halo of M31 (beyond $\sim$\,30\,kpc; \citealt{veljanoski14}), 
while leaving questions unanswered about the reliability of the previous findings in the inner regions.
This aspect may now be addressed through our comprehensive reanalysis of the M31 star cluster system, including
both high-quality imaging and a complete spectroscopic dataset (\citealt{C09} and \citealt{C11}, hereafter, Papers~I and II).
In particular, our age determinations allow us to securely differentiate old GCs from young-cluster and foreground-star contaminants.
The new data therefore
present an  opportunity to review the kinematics of the M31 GC system, and more generally to carry out a fresh global analysis of
its chemodynamical structure, along with a comparison to the MW.

This short paper is structured as follows.
In Section~\ref{sec:obs}, we discuss the observational data.
In Section~\ref{sec:spat}, we examine the spatial properties of the M31 and MW GC systems.
In Section~\ref{sec:kingroup}, we briefly analyze the kinematic properties of the three
major metallicity groups. 
A more detailed analysis of these trends will be left to future work.

\section{M31 Observations, Old and New}\label{sec:obs}

Our M31 GC sample is based on high signal-to-noise spectra from the MMT/Hectospec \citep{fab}, which provide
not only high-precision velocities ($\sim$\,6\,\kmsc), but also secure age and metallicity determinations from high signal-to-noise line indices.
 This key improvement allowed us to 
determine more precisely which clusters in our own working catalog (derived originally from the Revised Bologna Catalogue \cite{galleti}) were indeed old globular clusters.
Another advantage is that the vast majority of our metallicities are on the same system, allowing for a more coherent study of GC subpopulations 
\footnote{ the website \href{http://www.cfa.harvard.edu/oir/eg/m31clusters/M31\_Hectospec.html}{www.cfa.harvard.edu/oir/eg/m31clusters/M31\_Hectospec.html} contains images, 
spectra and other data on all of the known M31 star clusters.}.

Paper II determined metallicities using iron dominated Lick indices as measured on
Hectospec spectra, with the calibration supplied by similar measurements from the integrated
light of Galactic GCs. 
Those same spectra supplied the velocities used here, supplemented by
even more accurate velocities measured in the high dispersion spectra of \cite{strader}.
These spectroscopically determined metallicities were shown to correspond well to metallicities determined
from the color of the red giant branch (RGB) in {\it Hubble Space Telescope} produced color--magnitude diagrams (CMDs) in 22 MW GCs (since that publication, 
a further comparison with 45 more CMDs supplied by the Panchromatic Hubble Andromeda Treasury (PHAT) project 
has confirmed the agreement; \citealt{C15}). 
A few metallicities from Paper II have been refined (see table \ref{clusters}).
The formal uncertainties in the metallicities are
on average 0.15 dex; systematic uncertainties, while possibly present, should not affect our 
differential study here.
The median velocity uncertainty for this data set is 6 \kmsc, as derived from
clusters in common with
the \cite{strader} data, which had uncertainties of order 0.5 \kmsc.

We have further added a small number of new observations of previously known clusters to the collection,
as well as eleven clusters discovered in the PHAT study (nearly all in the bulge, and all low
mass). No new massive GCs (above $10^5$ M$_\sun$) were discovered in the NE part of the disk covered by the PHAT survey, and
thus we would not expect to find many such in the SW part of the disk.  The low mass PHAT bulge clusters, those with log $M/M_\sun < 4.5 $,  
are not included in the analysis here, because their inclusion  could bias the 
analysis  (the PHAT survey did not include the entire bulge). The complete sample
is listed in table \ref{clusters}.
This new spectroscopic and imaging work has also resulted in us classifying a handful
of other clusters as too young to be considered GCs, thus further revising some 
entries in Papers I and II. These are  B041-G103, B255D-D072, B258, B515, and B522, all of which have
ages less than 2 Gyr, based on their detailed spectra. These clusters are thus similar to
the large number of disk clusters with young ages mistakenly included previously by various authors as GCs, the latest of which was \cite{lee08}.

By combining the clusters identified as old in papers I and II (and excluding the few just
mentioned), the roughly 80 clusters of
the PAndAS survey, and the small number of bulge clusters from the PHAT survey, our best
estimate of the total number of known true GCs in M31  is currently 441. Previous
claims of more than 600 clusters were due to contamination by a large 
number of young disk clusters (\citealt{morrison04}, \citealt{puzia05} and \citealt{lee08}).
The optical disk of M31 is roughly 21 kpc in radius \citep[down to an {\it r}-band surface brightness of 24 mag arcs$^{-2}$;][]{kent1987},
within which there are 361 old GCs (we exclude eight clusters associated with NGC~205\footnote{ These are B009-G061, B011-G063, B317-G041, B328-G054, B330-G056, B331-G057, B333 and BH04}
).   Here are the sample sizes, where we express the number of clusters with log $M/M_\sun  > 4.5 $ in
parentheses. We have velocities for 344 (336) of these clusters, 
and metallicity estimates, either from the spectra, colors or CMDs 
for 346 (336), the vast majority coming from the 
uniform Hectospec study. The sample that has both velocities and metallicities
has 338 (332) members. Thus, our velocity and metallicity 
completeness within the 21 kpc radius is 94\%. Our
analysis here does not include the $\sim$ 80 halo clusters beyond 21 kpc. We refer
the reader to \cite{huxor14} and \cite{veljanoski14} for 
that discussion.
With this large and relatively uncontaminated sample, we hope to provide
a more definitive analysis of the kinematics of the M31 GCs as a function of metallicity grouping --
similar to the analysis of \citet{elson88} but with spectroscopic rather than photometric metallicities. 

We assume a distance  of 770 kpc \citep{freedman},  so that 1\arcmn=0.22 kpc.
We adopt an inclination of 77\degr, a minor axis position angle of $-52.3$\degr, a projected disk ellipticity 
of 0.7 in the outer parts (between 10 and 90\arcmn), an optical
disk scale length of 27\arcm (6~kpc; \citealt{kent1987}), and a projected bulge ellipticity of 0.3. We will occasionally 
use an $X$--$Y$ coordinate system for distance along the major and minor axes, with positive $X$ and $Y$
NE and NW of the center, respectively (e.g., M32 has negative $X$ and $Y$ coordinates).

\section{Spatial properties}\label{sec:spat}

In the MW, it is relatively straightforward to study the GC subpopulations and their properties, such as their spatial
distributions and kinematics, owing to their clear bimodality in metallicity---with peaks near [Fe/H]$=-1.6$ and $-0.6$.

For M31, we first examine the distribution of [Fe/H] versus galactocentric radius.
This is a standard approach (e.g., \citealt{barmby00}; Paper~II), 
but normally uses a simple circular radius---as would be appropriate for a spherical system.
If instead some subset of the GCs resides in an inclined, disklike distribution, then it
could be identified more clearly if disk coordinates were employed.
This is done by using disk isophote parameters from the previous Section to map the GCs to
the semi-major axis radius $R_a$, where
\begin{center}
$R_a = R~ [(1~ +~ (q^2~ -~ 1) ~cos^2 \phi)^{1/2}$]/q {\hskip 0.2truein}    (1) \\
\end{center}
and $R$ is the distance to the galaxy center (taken to be RA=0:42:44.3 Dec =+41:16:09.4),
q=0.3, the ratio of the assumed axes, and $\phi$ is the angle the cluster
makes with respect to the major axis and the center. The $R$ and $R_a$ values are also tabulated
in Table \ref{clusters}.

Figure \ref{feh_vs_r}  shows the results:
clusters more metal-rich than [Fe/H]\,$= -0.4$ are
not found outside of $R_a = 8$\,kpc.  
The same radius marks an apparent distinction in the density of GCs above and below [Fe/H]\,$ \sim -1.5$, 
with relatively few of the more metal-poor objects found at the smaller radii.
 These are the same metallicity divisions previously found in Paper~II,
and are also visible, though less clearly, when using bulge coordinates or simple
radial coordinates.

To better isolate these transitions, we calculate
the half-number radius for groups of GCs in different metallicity bins---i.e., the radius that divides a  group in half (again we leave out the smaller PHAT clusters).
For the overall sample of GCs within a limiting radius of R=21 kpc (with median [Fe/H]\,$=-1.0$), that half-number radius is $R=$\,4.2~kpc.  
For the 56 GCs with [Fe/H]\,$\geq -0.4$ (and median [Fe/H] $= -0.1$), 
it is much smaller, $R=$\,1.9~kpc.  
The 59 GCs with [Fe/H]\,$< -1.5$ (median [Fe/H]\,$=-1.9$) 
have a much larger half-number radius of $R=$\,6.3~kpc. 
Here we have experimented with different metallicity boundaries and thereby found these values where the
derived radius shows a transition---both for the disk radius and for the normal projected radius---noting
again that the uncertainty in metallicity for individual clusters is around 0.15~dex. 
The 223 GCs at intermediate metallicity (median [Fe/H]\,$=-1.0$) 
have a half-number radius of $R=$\,4.2~kpc, in between the values of the metal-rich and
metal-poor groups.

\section{Kinematics of Different Metallicity groups}\label{sec:kingroup}

We now investigate how the two metallicity dividing lines play out in the M31 cluster kinematics.
Figure \ref{abund_vs_vel} shows the GC velocities relative
to the mean M31 velocity ($-300$ \kmsc), with a sign inversion for
clusters on opposite sides of the rotating disk,  plotted against the cluster metallicity. 
That is, for GCs on the approaching SW side, $-$(V$-$V$_{\rm sys}$) is plotted, while on the NE, receding
side,  (V$-$V$_{\rm sys}$) is plotted.  In such
a plot, prograde velocities  will always be positive, while retrograde will be negative.
Viewed in this manner, in all three metallicity groups, there is net rotation 
in the same direction.
More than 90\% of clusters more metal-rich than [Fe/H]=$-0.4$ clearly have
prograde motion -- and no metal-rich cluster outside of 2 kpc has retrograde motion.
These facts imply disk kinematics for the most metal-rich clusters.
For  the intermediate-metallicity and metal-poor clusters, about 1/3 of the clusters
have retrograde motions. Excluding the inner 2 kpc objects, we find that 
again 1/3 of the intermediates have retrograde motion, while only 1/5 of the 
metal-poor clusters have
retrograde. It is no surprise that these groups are not  purely 
disk systems, but there is clearly some rotation, which we take up below.
\cite{allen06} and \cite{allen08} studied the orbits of 54 MW GCs, 32 of them more metal
poor than  [Fe/H] $=-0.8$ (the inflection point of the MW metallicity distribution) and found that half of those metal-poor clusters are on retrograde orbits, a much larger fraction than
found in M31.

Figure \ref{map} shows those three groups in a composite image, where the positions
in M31 are shown, along with their color-coded velocities. Our metallicity divisions
are sharp, and do not account for uncertainties in the metallicities, but the general
characteristics of this plot remain unchanged if we modify the group boundaries.
For the most metal-rich clusters, with [Fe/H]$>-0.4$, we find that all but one of
these 56 clusters are confined to the disk light distribution.  
The cluster concentration, noted above, is quite apparent, as is the systemic rotation,
and as we reported in \cite{morrison11},
the metal-rich clusters with $R<$ 2 kpc have apparent strong rotation, likely indicating a
response to a bar potential. Just as apparent, clusters with [Fe/H] $< -1.5$ are not concentrated, 
and are more spherically distributed, in projection.  Within 2 kpc, there are only 
7 clusters in our metal-poor group\footnote {These are  AP8925, B041D, B086-G148, B114-G175, 
B157-G212, B165-G218, and B264-NB19.}, whereas each of the other two groups has  roughly 
40. Even in this qualitative plot,
one can see that all three metallicity groups have some degree of rotation, though not
nearly as strong as reported in earlier papers cited above.

On the right side of this figure are plotted
the positions of MW clusters, in similar metallicity groups, but with different specific dividing lines.
The positions were determined from the galactic $XYZ$ coordinates in kpc provided in the \cite{harris96} MW GC
catalog from 2010, 
where we have projected those as follows: $X^{\prime} = ((X-{\rm R}_\sun)^2 + Y^2)^{1/2}$ and $Y^{\prime} = Z$, 
where R$_\sun =8$ kpc.  We grouped these in the following metallicity bins:  [Fe/H] $>-1.0$, $-1.5>$ [Fe/H] $>-1.0$,
and  [Fe/H] $<-1.5$. The first group includes clusters more metal-rich than  the saddle point in the overall
distribution, while the second and third groups divide the remainder at about the peak of the metal-poor
grouping.  As has been pointed out many times previously \citep[e.g., ][]{zinn}, the metal-rich group has
a large mean rotational velocity, a small line-of-sight dispersion, and a flattened spatial distribution.
The most metal-poor MW group again shows much less concentration than do the other two metallicity
bins. We cannot show velocities in the MW plot, but recall that the metal-poor MW clusters do not
show  bulk rotation  (\citealt{harris01}: $v/\sigma \sim 0.25$, where $\sigma$ is the group velocity
dispersion), unlike those in M31.

 To provide some numbers on the bulk rotation, 
we analyze the M31 GC radial velocities as a function of position angle
with respect to the minor axis of M31 (taken to be at a position angle of $-52.3$\degr).
We fit a simple sine function to these data using non-linear least squares 
(cf.\ \citealt{sharples}) \\
\begin{center}
$V(\theta) = V_0 + K $sin $(\theta - \theta_0)$ {\hskip 0.4truein} (2) \\
\end{center}
where $V(\theta)$ is the cluster observed radial velocity at position angle $\theta$, 
$V_0$ is the group mean velocity, $K$ is the 
amplitude of rotation, and $\theta_0$ is the position angle of the axis of rotation\footnote{We have also experimented with
fits that include azimuthal flattening, equivalent to tilted-ring models (e.g., \citealt{foster}), 
and find similar results,
though with somewhat stronger rotation.}.
We fit each of our three defined metallicity groups separately, and derived the dispersions
about those fits (see Figure~\ref{rotation}). The results are 
presented in Table  \ref{rotation_table}, where we have further broken up the metal-rich
sample into bulge and non-bulge clusters (those with $R >$ 2 kpc). The non-bulge metal-rich group
has a bulk rotation rate of around 160 \kmsc, with a moderate dispersion of 80 \kms\ ($v/\sigma \sim 2.0$).
The intermediate-metallicity
group has a rotation of 53 \kmsc, and a dispersion of 141 \kmsc\ ($v/\sigma \sim 0.4$), 
while the metal-poor group has a somewhat
higher rotation of 90 \kms and a similar dispersion of 154 \kmsc\ ($v/\sigma \sim 0.6$). 
The detected bulk rotation rates are significantly non-zero for the intermediate and metal-poor groups,
$4.0\sigma$ and $3.8\sigma$, respectively (from Table  \ref{rotation_table}, where $\sigma$ here is
the uncertainty in the derived parameter) Similar results are found if
we fix the mean velocity of all the subgroups to be equal to that of the full group.

For the metal-rich group, we can also compare the observed radial
velocities with those expected for the thin disk at those positions, 
using the HI$+$HII-based rotation model of
\cite{kent}, and derive a dispersion from the differences.  
The rotation model uses a major axis curve that rises linearly to 
250 \kms out to a radius of 6.5 kpc, and is flat outside of that. Projected
velocities throughout the disk are then given by V=V$_{rot}$ \,cos $\theta $\,sin $i$,
where V$_{rot}$ is the major axis rotation at the distance corresponding to
the location of the GC, $\theta$ is the angle with respect to the major
axis, and $i$ is again the inclination.
Again, we find a dispersion about the thin disk rotation of 80 \kms for the metal-rich group. 
A similar result is derived if we use the actual M31 HI velocity map of C.\ Lee and A.K.\ Leroy (priv.comm.),
though not all clusters have HI detected at their location. It is to be expected that
these old clusters would have higher dispersions than a thin disk of HI gas.

The dispersion in 140 HII region velocities about the disk model as derived from the data presented in
 \cite{sanders} is about 50 \kmsc. From the data in Paper I we can also derive the
dispersion of the diffuse disk gas (from 60 pointings), and that of 60 clusters younger than 1 Gyr. 
Those values are 39 and 41 \kms respectively. 
Another comparison can be made to results from the
Spectroscopic and Panchromatic Landscape of Andromeda's Stellar Halo Survey
\citep{dorman12,dorman13,dorman15}.
Here the kinematics of individual RGB stars have been analyzed, and separated into disk and spheroid components.
In the same radial range as the metal-rich GCs, the disk stars have
velocity dispersions of $\sim$~90--130~\kmsc, 
while the spheroid has a dispersion of $\sim$~120--160~\kmsc.
Thus, the metal-rich GCs appear to track the galaxy disk rather than the spheroid (or extended bulge).

The two more metal-poor GC metallicity groups also have mild systemic prograde rotation, though with larger dispersions.
Their $\sim$~50--100\kms rotation  is lower than the $\sim$120--200\kms values
that have been circulating in the literature for the past decade.
Our revised value also matches up nicely with the outer halo
rotation of $\sim$~80\kmsc, which had previously shown a peculiar disconnect with the inner halo
(see figure~2 from \citealt{veljanoski13} and figure~7 from \citealt{veljanoski14}),
owing to the inner GC sample being contaminated with very young clusters.
Our updated summary of all available M31 GC velocities, out to $\sim$~100~kpc, is shown in Figure~\ref{veljanoski}.

Relating back to the field star components again,
the intermediate-metallicity GCs are kinematically similar to the RGB extended spheroid stars, but have a lower metallicity
(median [Fe/H] $\sim -1.0$ versus $-0.5$; \citealt{gilbert14}). 
Instead, they may be associated with the metal-poor ``halo'' detected through RGB stars
\citep{gilbert12,ibata}.
The most metal-poor GC group has higher rotation than the intermediate group, and 
is spatially the least concentrated of the three groups, 
by a factor of at least five within a radius of 2 kpc, as calculated from the positions listed in
Table \ref{clusters}.
Again, it may be associated with the RGB halo, and we note that the presence of a metal-poor stellar halo in M31,
and in many other galaxies, was evident from the GCs, long before resolved field-star studies were feasible.

There is some asymmetry in the bulk rotation patterns of the GCs to comment on.
For instance, the  minimum observed radial velocity of unresolved optical light
on the approaching side (SW, negative $X$, right side of the M31 image in Fig.~\ref{map}) is
about $-250$\kms (relative to systemic; from data reported in Paper I).
The maximum velocity on the receding side (NE, positive $X$, left side) is
about $+350$\kms.  There are no clusters on the approaching side with velocities 
greater than $+200$\kms which are farther than
1 kpc from the center (i.e., with velocities that differ from the local
disk velocity by $+450$\kms), but there are four clusters
on the receding side with velocities less than $-200$\kms ($-450$\kms with
respect to the local disk), three of
them projected on the disk. These are V129-BA4, B213-B264, and B173-G224. There is
nothing else unusual about these clusters, but again perhaps more
detailed analysis is warranted to see if they were deposited by the Giant Southern Stream or another
stream.

One metal-rich cluster, B094-G156, has an observed radial velocity
that is $-202$\kms different from the local disk velocity (though prograde), 
making its velocity difference similar to the bar-influenced clusters at smaller radii, even
though it is much farther from the center (4 kpc along the minor axis) than the other such clusters reported 
in \cite{morrison11}, who considered clusters within a radius of 2 kpc.

One final comment to make regards the four metal-rich clusters with [Fe/H] $\sim -0.4$ 
that lie outside
the optical disk, and at radii larger than 10 kpc, more than twice as far as
the next farthest metal-rich cluster. These are B339-G077, 
B379-G312, B398-G341, and B407-G352.  
B407-G352 has been suggested as the remnant nucleus
of the Giant Southern Stream \citep{perina09}. Perhaps the other three outer, high metallicity
clusters bear further investigation with that topic in mind.

\section{Summary and Discussion}

We now return to comparisons of the MW and M31 GC systems (Section~\ref{intro}).
First, as is well known, M31 has more than twice the total number of GCs as the MW. 
Within the optical disks of both galaxies (using 21 kpc for both),
there are 361 GCs in M31 and 129 in the MW (using \citealt{harris96} for the MW numbers).
Larger numbers for M31 reported previously  are contaminated by the inclusion of young disk clusters. 

\begin{figure}[ht]
\includegraphics[width=3.3in]{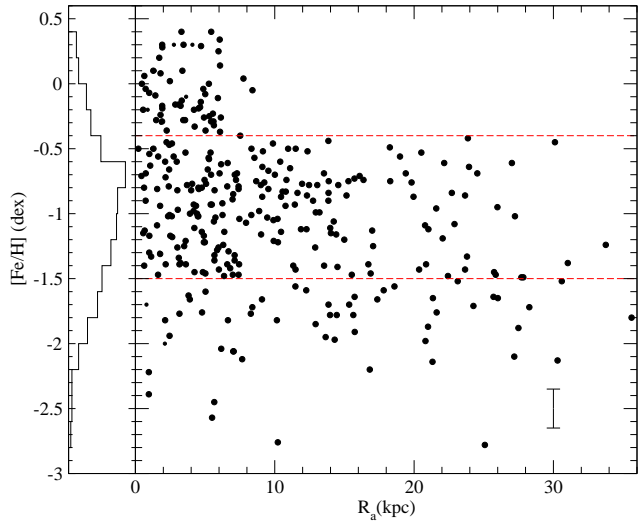} 
\caption{Distribution of M31 GC iron metallicity with semi-major axis radius, projected elliptically to the disk.
In this plot, three metallicity groups emerge (with divisions at [Fe/H]\,$\sim -0.4$ and $-1.5$,
shown as dashed lines), 
based on their relative densities inside and outside a disk radius of $\sim$\,8\,kpc. 
 Note that eight clusters around NGC~205 (at $R_a=24-28$\,kpc) 
have been removed from this plot, and there are also
another 115 GCs (mostly from the PAndAS survey) that either have
no known metallicity or extend off the plot to larger radii. 
Ten clusters with log mass in solar units less than 4.5 are shown as smaller dots (mostly at small radii).
As a result, there are 326 clusters shown here.
The median metallicity uncertainty of $\pm\,0.15$~dex is shown by error-bars at lower right.
The binned metallicity distribution among all of clusters (including those off the radius scale)
is shown at the left, reiterating the finding
in \cite{C11} that the distribution is not simply bimodal.
\label{feh_vs_r}}
\end{figure}

\begin{figure}[ht]
\includegraphics[width=3.5in,clip=true]{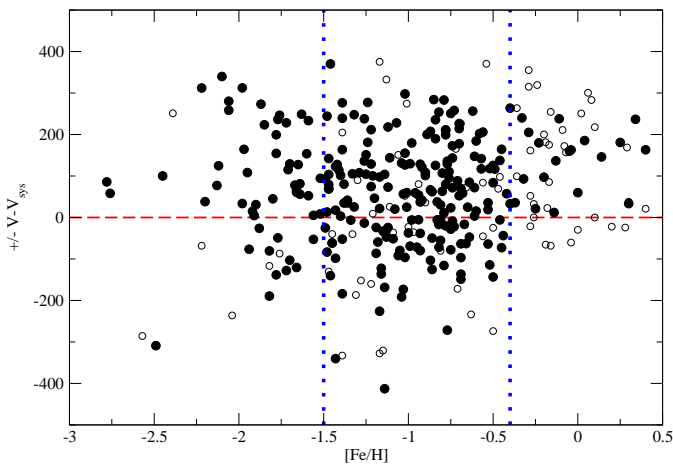}
\caption{A second diagnostic plot outlining M31 GC metallicity groups.
Here, velocity with respect to systemic velocity, with a sign
inversion on opposite sides of the rotating disk, is plotted against
[Fe/H].  Thus, objects with positive ordinate values are rotating prograde;
retrograde velocities result in negative ordinate values.  Clusters closer than
2 kpc to the center are shown as open circles; those farther out are shown
as filled circles. This plot again suggests that
clusters can be divided at [Fe/H]$=-0.4$ and  [Fe/H]$=-1.5$, shown by the dotted vertical lines.  Nearly all
clusters more metal-rich than  [Fe/H]$=-0.4$ have prograde motions.
Lower metallicity clusters have roughly twice as many prograde as retrograde
clusters, indicating systemic rotation for them. 
\label{abund_vs_vel}}
\end{figure}

Second, while
the metallicity distribution for the MW is clearly bimodal, that of M31 is not 
a simple superposition of Gaussians. In this paper, we have suggested three components: a very
metal-rich group, the dominant intermediate-metallicity group, and a metal-poor group. These
groups are most easily defined by their differing two dimensional spatial distributions, as shown in Figure \ref{feh_vs_r}.

Third, outside of the central bulge regions, the metal-rich group (20 clusters) has convincing disk kinematics
(both in rotation and velocity dispersion),
has a spatial distribution like the optical M31 disk, and is much more concentrated than the other two GC groups.
The lower-metallicity groups have weak but significant prograde rotation. 
The metal-poor group is the least concentrated of the three groups, down by a factor of at least five.
These results obviate the need to invoke a major merger \citep[see also][]{vel2016}, and they bring the
properties of the M31 GC system into closer alignment with the MW.
Although the numbers of GCs as a function of metallicity are different for the two galaxies,
their spatial and kinematical trends with metallicity are fairly similar
-- as was previously emphasized by \citet{elson88}.

The clear disk-like properties of the M31 metal-rich GCs provide an important window into the
formation mechanisms of both galaxies and star clusters.
In the Milky Way, the difficulties in observing the metal-rich GC sub-system leads to ambiguity in classifying
them as bulge or thick-disk objects. 
Consequently, an association between bulges and metal-rich GCs is often assumed,
along with gas-rich major mergers as the mechanism that formed these GCs
(e.g., \citealt{ashman,li14}).
This picture has been tested recently in lenticular galaxies through detailed comparisons of metal-rich GCs to
the bulge and disk field-star components
\citep{forbes12,cortesi}, with mixed results.
However, the case of M31 highlights an alternative to the merger scenario:
 that a population of GCs formed in-situ, within giant star-forming clumps in turbulent galactic disks
at high redshifts  \citep{shapiro,kruijssen}.

Although this disk mode of GCs is intriguing, the very dominant subpopulations in M31
are the lower-metallicity GCs.  It seems likely that these objects were brought in by the accretion of
satellite galaxies (minor mergers). The populous GC system of M31, relative to the MW,
would then reflect a more active accretion history, and the radial gradient of GC metallicities
could arise through correlations of both metallicity and dynamical friction with satellite galaxy mass
(e.g., \citealt{amorisco}).

The chemodynamical structure of M31, and its assembly history, could be clarified in the future
by comparing the GCs to the stars in more detail.
Analysis of additional spiral galaxies and their GC systems may also reveal how
pervasive the different modes of GC formation are, and how representative M31 and the MW are of other galaxies.

\acknowledgments
Important
discussions were had with Jay Strader, Anil Seth, Ricardo Schiavon, Heather Morrison, and Claire Dorman.
We thank Adam Leroy and Cheoljong Lee for the use of the M31 HI map.

\clearpage

\clearpage

\begin{figure}[hb]
\includegraphics[width=7.0in]{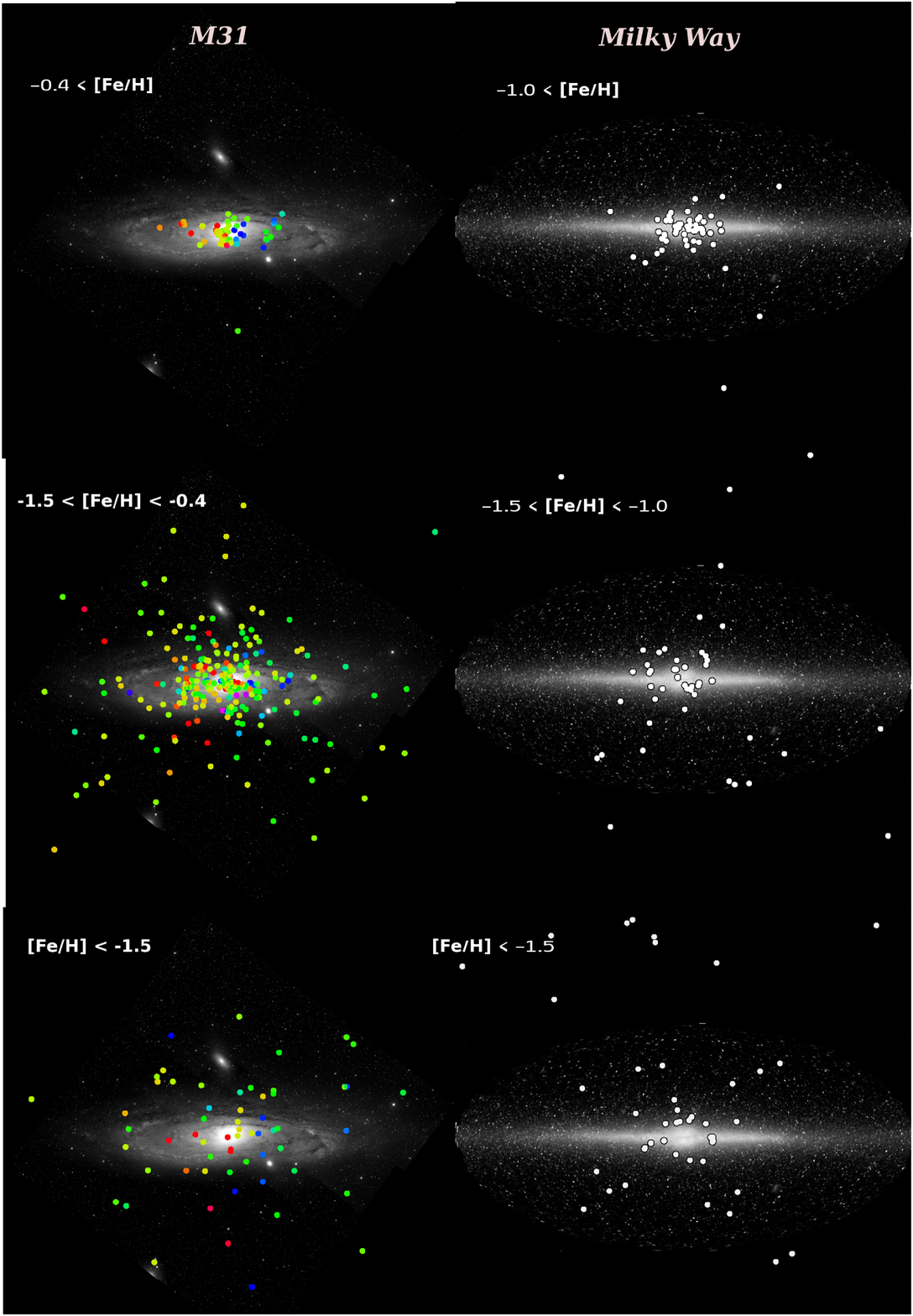}

\end{figure}
\clearpage

\begin{figure}[hb]
\caption{Left panels: Maps of the locations of different [Fe/H] groups for M31 GCs, 
color coded by velocity. In the upper plot are the
most metal-rich, [Fe/H]$> -0.4$. The middle plot shows clusters with $-0.4<$[Fe/H]$<-1.4$,
while the lower panel contains the most metal-poor group. 
The background image is from the DSS, where the field size is 50 kpc. 
The color coding is with respect to the mean
M31 velocity, and is as follows: violet$=-375$,  blue$=-275$, green$=-100$, yellow$=+50$, orange$=+150$ and
red $=+225$ \kms.
The most metal-poor group is much less concentrated than the metal-rich group.
 The outlying cluster in the metal-rich group is B379-G312, discussed in the text.
Right panels:  similar plots for MW
clusters, where the projection of galactic $XYZ$ onto a plane is described in the text.
At the top are the most metal-rich clusters,
with  [Fe/H]$> -1.0$, which represents the clusters of the metal-rich peak in the overall
metallicity distribution.
The middle panel shows clusters with $-1.0<$[Fe/H]$<-1.5$, and the bottom has
clusters with [Fe/H]$< -1.5$. The dividing line here is at the peak of the metal-poor group in the MW.
The background image is the DIRBE/COBE image of the MW (courtesy NASA \& E. Wright). We arbitrarily
set the MW image to be 40 kpc in diameter for this display, similar to the size of the
image shown of M31. 
Like the situation for M31,
the metal-poor MW clusters show less concentration than the metal-rich group.
As general information, this plot also shows that M31 has 361 known GCs
within a projected radius of 21 kpc, while the MW has just 129.
\label{map}}
\end{figure}

\clearpage
\begin{figure}[ht]
\includegraphics[width=3.3in]{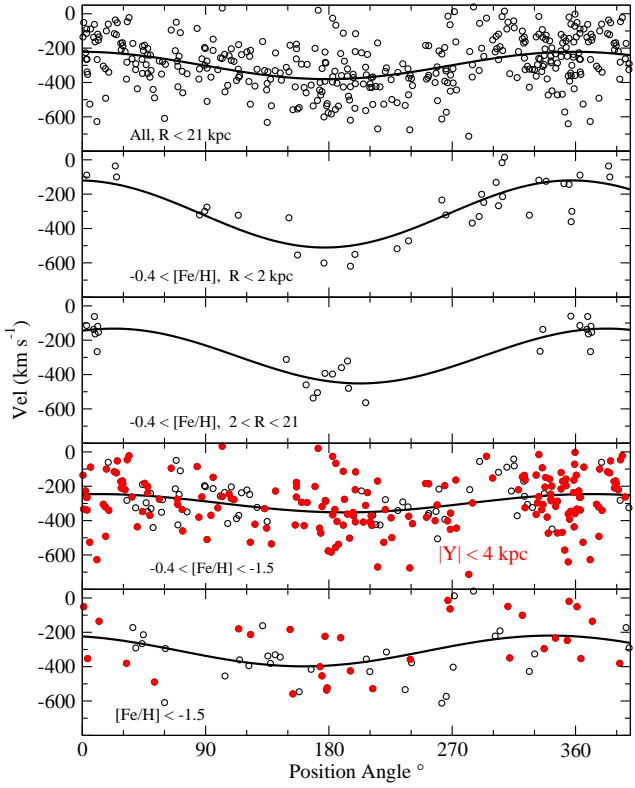}
\caption{
Radial velocity plotted as a function of position angle with respect to the minor axis,
in metallicity and radius bins. 
Curves show model fits for mean velocity versus position angle.
All data refer to clusters with radius $< 21$ kpc. At the top are plotted
all such clusters, demonstrating that the entire sample has bulk rotation at the level of 80\kms,
 shown as the continuous line. 
Table  \ref{rotation_table} lists the derived values for all the sub-groups shown. In the lower two panels, 
filled red circles refer to clusters projected to within 4 kpc of the stellar disk, to search more
closely for disk-like rotation in these two more metal-poor groups.
\label{rotation}}
\end{figure}
\clearpage
\begin{figure}[ht]
\includegraphics[width=3.3in]{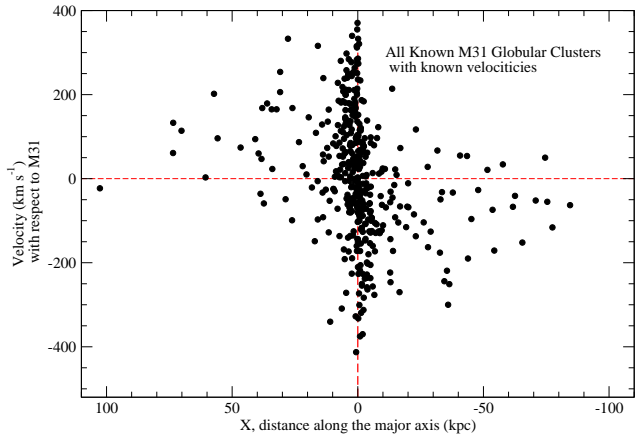}
\caption{
Radial velocity plotted as a function of projected distance along the major axis ({\it X}), for
all M31 GCs without radius restriction, with measured velocities  (421 out of the full 441 M31 GCs that we have collected), including the outer clusters discussed in \cite{veljanoski14}. 
The outer clusters (those with  $ X >$ 21) do show the same sense of rotation as the inner ones as reported, but the version
of the inner data shown in that paper was contaminated by young disk clusters. Dashed lines show the zero levels for both axes.
\label{veljanoski}}
\end{figure}


\clearpage

\LongTables

\pagestyle{empty}
\begin{deluxetable}{lllrrrrrrrr}
\tabletypesize{\scriptsize}
\setlength{\tabcolsep}{0.1in} 
\tablecolumns{11}
\tablewidth{6.4truein}
\tablecaption{All Known Globular Clusters in Our M31 Sample\label{clusters}} 
\tablehead{\colhead{Object} &\colhead{RA}  &\colhead{Dec}  &\colhead{Velocity}  &\colhead{Ref\tablenotemark{a}}   &\colhead{[Fe/H]\tablenotemark{b}}  &\colhead{Ref\tablenotemark{c}}   &\colhead{Age\tablenotemark{d}} &\colhead{Log M\tablenotemark{e}}  &\colhead{$R$\tablenotemark{f}}  &\colhead{$R_a$\tablenotemark{g}} \\
\colhead{} & \multicolumn{2}{c}{J2000}    &\colhead{\kms}  &\colhead{}  &\colhead{} &\colhead{} &\colhead{Gyr}  &\colhead{M$_\sun$}  &\colhead{kpc} &\colhead{kpc}\\
}

\startdata
PAndAS-01&	23:57:12.0&	43:33:08&	$-$333.0$\pm	10.0$&	hx&		&	&	(14)&	5.3&	117.9&	374.1	\\
PAndAS-02&	23:57:55.6&	41:46:49&	$-$266.0$\pm	10.0$&	hx&		&	&	(14)&	5.1&	113.7&	331.6	\\
PAndAS-04&	0:04:42.9&	47:21:42&	$-$397.0$\pm	10.0$&	hx&		&	&	(14)&	5.2&	123.6&	409.0	\\
shortened - see journal for full table

\enddata
\tablenotetext{a}{References for velocities:  b=\cite{barmby00}; co=\cite{colucci}; p=\cite{perrett02}; hs2=Hectospec from Paper II,  hs7=Hectospec from here; he=Hectochelle from \cite{strader};  hx=\cite{huxor14} and references therein; rbc=\cite{galleti09}.}
\tablenotetext{b}{Minimum uncertainties set to 0.1.}
\tablenotetext{c} {References for [Fe/H]: hs= Hectospec from Paper II or here (``hs7''); c= derived from PHAT colors (Caldwell et al. 2016 in prep);  co=\cite{colucci}; ); m06=CMD value from  \cite{mackey06};  m07=CMD value from \cite{mackey07} ;  m13=CMD value from \cite{mackey13} ;  pa=CMD value from \cite{huxor14}; ph=CMD value from PHAT data, (Caldwell et al. 2016 in prep); p11=\cite{perina11};  r05=CMD value from \cite{rich}. Blank where no Hectospec spectroscopy or HST CMD exists, and PHAT
photometry is not available.}
\tablenotetext{d}{From paper II. Values in parentheses assigned where no precise age was determined.}
\tablenotetext{e}{Log of total mass, from photometry in paper II or here.}
\tablenotetext{f}{Radial distance to center of M31.}
\tablenotetext{g}{Defined in the text.}
\end{deluxetable}

\clearpage

\begin{deluxetable}{lrrrrr}
\tablecolumns{6}
\tablewidth{0pc}
\tablecaption{Rotation of Different Metallicity Groups\tablenotemark{a}\label{rotation_table} } 
\tablehead{\colhead{Group} &\colhead{N} & \colhead{V$_0$}  & \colhead{K} &   \colhead{$\theta_0$\tablenotemark{b}} &\colhead{$\sigma $}  \\
\colhead{} & \colhead{} & \colhead{\kms} & \colhead{\kms} & \colhead{Deg.} & \colhead{\kms} 
}
\startdata
All, R$<21$ kpc, Log $M/M_\sun < 4.5$    &332&$-301\pm 8$ & 80$\pm 10$ & $89\pm 9$ & 140\\
$-0.4<$[Fe/H]              &54& $-302\pm 14$ & $168\pm 16$ & $89\pm 9$ & 104 \\
$-0.4<$[Fe/H], R$<2$  &33& $-315\pm 24$ & $195\pm 31$& $87\pm 9$& 120 \\
$-0.4<$[Fe/H], 2$<$R$<$21 &20& $-292\pm 18$ & $160\pm 35$& $113\pm 25$& 80 \\
$-1.5<$[Fe/H]$<-0.4$              &221& $-299\pm 10$& 53$\pm 13$& $101\pm 16$& 141 \\
$-1.5<$[Fe/H]$<-0.4$, $|$Y$|< 4$ kpc             &162& $-304\pm 12$& 53$\pm 17$& $108\pm 22$& 152 \\
$[$Fe/H$]<-1.5$      &57& $-309\pm 20$ & $90\pm 23$& $71\pm 24$& 154 \\
$[$Fe/H$]<-1.5$, $|$Y$|<4$      &27& $-285\pm 29$ & $112\pm 42$& $51\pm 31$& 154 \\

\enddata
\tablenotetext{a}{Uncertainties were determined by a bootstrap method.}
\tablenotetext{b}{90\degr \ is the photometric minor axis.}
\end{deluxetable}


\clearpage
\begin{thebibliography}{}



\bibitem[Allen et al.(2006)]{allen06} Allen, C., Moreno, E., \& Pichardo, B.\ 2006, \apj, 652, 1150 
\bibitem[Allen et al.(2008)]{allen08} Allen, C., Moreno, E., \& Pichardo, B.\ 2008, \apj, 674, 237 
\bibitem[Amorisco(2016)]{amorisco} Amorisco, N.~C.\ 2016, \mnras, submitted, arXiv:1511.08806
\bibitem[Ashman \& Zepf(1992)]{ashman} Ashman, K.~M., \& Zepf, S.~E.\ 1992, \apj, 384, 50 



\bibitem[Barmby et al.(2000)]{barmby00} Barmby, P., Huchra, J.~P., Brodie, J.~P., et al.\ 2000, \aj, 119, 727 
\bibitem[Beasley et al.(2004)]{beasley04} Beasley, M.~A., Brodie, J.~P., Strader, J., et al.\ 2004, \aj, 128, 1623 
\bibitem[Bekki(2010)]{bekki10} Bekki, K.\ 2010, \mnras, 401, L58 

\bibitem[Caldwell et al.(2009)]{C09} Caldwell, N., Harding, P., Morrison, H., Rose, J.~A., Schiavon, R., \& Kriessler, J.\ 2009, \aj, 137, 94 
\bibitem[Caldwell et al.(2011)]{C11} Caldwell, N., Schiavon, R., Morrison, H., Rose, J.~A., \& Harding, P.\ 2011, \aj, 141, 61 
\bibitem[Caldwell in prep.(2016)]{C15} Caldwell, N., et al.\ in prep.\ 2016

\bibitem[Carollo et al.(2007)]{carollo07} Carollo, D., Beers, T.~C., Lee, Y.~S., et al.\ 2007, \nat, 450, 1020 
\bibitem[Cezario et 
al.(2013)]{cz13} Cezario, E., Coelho, P.~R.~T., Alves-Brito, A., Forbes, D.~A., \& Brodie, J.~P.\ 2013, \aap, 549, A60 
\bibitem[Colucci et al.(2014)]{colucci} Colucci, J.~E.,  Bernstein, R.~A., \& Cohen, J.~G.\ 2014, \apj, 797, 116 
\bibitem[Cortesi et al.(2016)]{cortesi} Cortesi, A.,  Chies-Santos, A.~L., Pota, V., et al.\ 2016, \mnras, 456, 2611

\bibitem[Deason et al.(2011)]{deason11} Deason, A.~J., Belokurov, V., \& Evans, N.~W.\ 2011, \mnras, 411, 1480 
\bibitem[Dorman et al.(2012)]{dorman12} Dorman, C.~E., Guhathakurta, P., Fardal, M.~A., et al.\ 2012, \apj, 752, 147 
\bibitem[Dorman et al.(2013)]{dorman13} Dorman, C.~E., Widrow, L.~M., Guhathakurta, P., et al.\ 2013, \apj, 779, 103
\bibitem[Dorman et al.(2015)]{dorman15} Dorman, C.~E.,  Guhathakurta, P., Seth, A.~C., et al.\ 2015, \apj, 803, 24 

\bibitem[Elson \& Walterbos(1988)]{elson88} Elson, R.~A., \& Walterbos, R.~A.~M.\ 1988, \apj, 333, 594 

\bibitem[Fabricant et al.(2005)]{fab} Fabricant, D., et  al.\ 2005, \pasp, 117, 1411 

\bibitem[Forbes \& Bridges(2010)]{forbes10} Forbes, D.~A., \& Bridges, T.\ 2010, \mnras, 404, 1203 
\bibitem[Forbes et al.(2012)]{forbes12} Forbes, D.~A., Cortesi, A., Pota, V., et al.\ 2012, \mnras, 426, 975 
\bibitem[Foster et al.(2011)]{foster} Foster, C., Spitler,  L.~R., Romanowsky, A.~J., et al.\ 2011, \mnras, 415, 3393 
\bibitem[Freedman \& Madore(1990)]{freedman} Freedman, W.~L., \& Madore, B.~F.\ 1990, \apj, 365, 186 
\bibitem[Freeman(1983)]{freeman83} Freeman, K.~C.\ 1983, Internal Kinematics and Dynamics of Galaxies, 100, 359 

\bibitem[Galleti et al.(2007)]{galleti} Galleti, S.,  Bellazzini, M., Federici, L., Buzzoni, A., \& Fusi Pecci, F.\ 2007, \aap, 471, 127 
\bibitem[Galleti et al.(2009)]{galleti09} Galleti, S., Bellazzini, M., Buzzoni, A., Federici, L., \& Fusi Pecci, F.\ 2009, \aap, 508, 1285 
\bibitem[Gilbert et al.(2012)]{gilbert12} Gilbert, K.~M., Guhathakurta, P., Beaton, R.~L., et al.\ 2012, \apj, 760, 76 
\bibitem[Gilbert et al.(2014)]{gilbert14} Gilbert, K.~M., Kalirai, J.~S., Guhathakurta, P., et al.\ 2014, \apj, 796, 76


\bibitem[Harris(1996)]{harris96} Harris, W.~E.\ 1996, VizieR Online Data Catalog, 7195, 0 
\bibitem[Harris(2001)]{harris01} Harris, W.~E.\ 2001, Saas-Fee  Advanced Course 28: Star Clusters, 223 
\bibitem[Huchra et al.(1982)]{huchra82} Huchra, J., Stauffer, J., \& van Speybroeck, L.\ 1982, \apjl, 259, L57 
\bibitem[Huchra et al.(1991)]{huchra91} Huchra, J.~P., Brodie,  J.~P., \& Kent, S.~M.\ 1991, \apj, 370, 495 
\bibitem[Huxor et al.(2014)]{huxor14} Huxor, A.~P., Mackey, A.~D., Ferguson, A.~M.~N., et al.\ 2014, \mnras, 442, 2165 


\bibitem[Ibata et al.(2014)]{ibata} Ibata, R.~A., Lewis, G.~F., McConnachie, A.~W., et al.\ 2014, \apj, 780, 128 



\bibitem[Keller et al.(2012)]{keller12} Keller, S.~C., Mackey, D., \& Da Costa, G.~S.\ 2012, \apj, 744, 57 
\bibitem[Kent(1987)]{kent1987} Kent, S.~M.\ 1987, \aj, 94, 306 
\bibitem[Kent(1989)]{kent} Kent, S.~M.\ 1989, \pasp, 101, 489 
\bibitem[Kruijssen(2015)]{kruijssen} Kruijssen, J.~M.~D.\ 2015, \mnras, 454, 1658 

\bibitem[Lee et al.(2008)]{lee08} Lee, M.~G., Hwang, H.~S.,  Kim, S.~C., Park, H.~S., Geisler, D., Sarajedini, A.,  \& Harris, W.~E.\ 2008, \apj, 674, 886 
\bibitem[Li \& Gnedin(2014)]{li14} Li, H., \& Gnedin, O.~Y.\ 2014, \apj, 796, 10 

\bibitem[Mackey \& Gilmore(2004)]{mackey04} Mackey, A.~D., \& Gilmore, G.~F.\ 2004, \mnras, 355, 504 
\bibitem[Mackey et al.(2007)]{mackey07} Mackey, A.~D., Huxor, 
A., Ferguson, A.~M.~N., et al.\ 2007, \apjl, 655, L85 
\bibitem[Mackey et al.(2006)]{mackey06} Mackey, A.~D., Huxor, 
A., Ferguson, A.~M.~N., et al.\ 2006, \apjl, 653, L105 
\bibitem[Mackey et al.(2013)]{mackey13} Mackey, A.~D., Huxor, 
A.~P., Ferguson, A.~M.~N., et al.\ 2013, \mnras, 429, 281 
\bibitem[McCarthy et al.(2012)]{mccarthy} McCarthy, I.~G., Font, A.~S., Crain, R.~A., et al.\ 2012, \mnras, 420, 2245
\bibitem[McConnachie et al.(2009)]{mcconnachie09} McConnachie, A.~W., Irwin, M.~J., Ibata, R.~A., et al.\ 2009, \nat, 461, 66 
\bibitem[Morrison et al.(2004)]{morrison04} Morrison, H.~L.,  Harding, P., Perrett, K., \& Hurley-Keller, D.\ 2004, \apj, 603, 87 
\bibitem[Morrison et al.(2009)]{morrison09} Morrison, H.~L., Helmi, A., Sun, J., et al.\ 2009, \apj, 694, 130 
\bibitem[Morrison et al.(2011)]{morrison11} Morrison, H., Caldwell, N., Schiavon, R.~P., et al.\ 2011, \apjl, 726, L9 



\bibitem[Perrett et al.(2002)]{perrett02} Perrett, K.~M., Bridges, T.~J., Hanes, D.~A., Irwin, M.~J., Brodie, J.~P., Carter, D., Huchra, J.~P., \& Watson, F.~G.\ 2002, \aj, 123, 2490
\bibitem[Perina et al.(2009)]{perina09} Perina, S., Federici, L., Bellazzini, M., et al.\ 2009, \aap, 507, 1375 
\bibitem[Perina et 
al.(2011)]{perina11} Perina, S., Galleti, S., Fusi Pecci, F., et al.\ 2011, \aap, 531, A155 
\bibitem[Puzia et al.(2005)]{puzia05} Puzia, T.~H., Perrett, K.~M., \& Bridges, T.~J.\ 2005, \aap, 434, 909

\bibitem[Rich et al.(2005)]{rich} Rich, R.~M., Corsi, C.~E., 
Cacciari, C., et al.\ 2005, \aj, 129, 2670 



\bibitem[Searle \& Zinn(1978)]{SZ1978} Searle, L., \& Zinn, R.\ 1978, \apj, 225, 357 
\bibitem[Sanders et al.(2012)]{sanders} Sanders, N.~E.,  Caldwell, N., McDowell, J., \& Harding, P.\ 2012, \apj, 758, 133 
\bibitem[Shapiro et al.(2010)]{shapiro} Shapiro, K.~L., Genzel, R., F\"orster Schreiber, N.~M.\ 2010, \mnras, 403, L36 
\bibitem[Sharples et al.(1998)]{sharples} Sharples, R.~M., Zepf, S.~E., Bridges, T.~J., et al.\ 1998, \aj, 115, 2337 
\bibitem[Strader et al.(2011)]{strader} Strader, J., Caldwell,  N., \& Seth, A.~C.\ 2011, \aj, 142, 8 

\bibitem[Veljanoski et al.(2013)]{veljanoski13} Veljanoski, J.,  Ferguson, A.~M.~N., Mackey, A.~D., et al.\ 2013, \apjl, 768, L33 
\bibitem[Veljanoski et al.(2014)]{veljanoski14} Veljanoski, J., Mackey, A.~D., Ferguson, A.~M.~N., et al.\ 2014, \mnras, 442, 2929
\bibitem[Veljanoski \& Helmi(2016)]{vel2016} Veljanoski, J., \& Helmi, A. \ 2016, \aap, submitted, arXiv:1602.04018

\bibitem[Zinn(1985)]{zinn85} Zinn, R.\ 1985, \apj, 293, 424 
\bibitem[Zinn(1996)]{zinn} Zinn, R.\ 1996, Formation of the Galactic Halo...Inside and Out, 92, 211 





\end{thebibliography}
\end{document}